\documentclass{article}
\usepackage[utf8]{inputenc}
\usepackage{epsfig}
\usepackage{authblk}
\usepackage{xcolor}
\usepackage{overpic}
\usepackage{amssymb}
\usepackage{amsmath}

\usepackage[margin=1in]{geometry}
\usepackage{multirow}
\usepackage{soul}
\usepackage[colorlinks=true, linkcolor=blue, filecolor=blue, citecolor=blue]{hyperref}
\usepackage[square, comma, sort&compress,numbers]{natbib}
\usepackage{lineno}
\usepackage{comment}
\usepackage{hyperref}
\usepackage{setspace}

\title{Explaining the physics of transfer learning a data-driven subgrid-scale closure to a different turbulent flow}

\author{Adam Subel$^1$\thanks{Current affiliation: Courant Institute of Mathematical Sciences, New York University, New York City 10012 NY} , Yifei Guan$^1$, Ashesh Chattopadhyay$^1$, and Pedram Hassanzadeh$^{1,2}$\thanks{pedram@rice.edu}}
\affil{\normalsize $^1$Department of Mechanical Engineering, Rice University, Houston 77005 TX}
\affil{\normalsize $^2$Department of Earth, Environmental and Planetary Sciences, Rice University, Houston 77005 TX}
\date{}

\begin{document}

\maketitle

\begin{abstract}
Transfer learning (TL) is becoming a powerful tool in scientific applications of neural networks (NNs), such as weather/climate prediction and turbulence modeling. TL enables out-of-distribution generalization (e.g., extrapolation in parameters) and effective blending of disparate training sets (e.g., simulations and observations). In TL, selected layers of a NN, already trained for a base system, are re-trained using a small dataset from a target system. For effective TL, we need to know 1) what are the best layers to re-train? and 2) what physics are learned during TL? Here, we present novel analyses and a new framework to address (1)-(2) for a broad range of multi-scale, nonlinear systems. Our approach combines spectral analyses of the systems' data with spectral analyses of convolutional NN's activations and kernels, explaining the inner-workings of TL in terms of the system’s nonlinear physics. Using subgrid-scale modeling of several setups of 2D turbulence as test cases, we show that the learned kernels are combinations of low-, band-, and high-pass filters, and that TL learns new filters whose nature is consistent with the spectral differences of base and target systems. We also find the shallowest layers are the best to re-train in these cases, which is against the common wisdom guiding TL in machine learning literature. Our framework identifies the best layer(s) to re-train beforehand, based on physics and NN theory. Together, these analyses explain the physics learned in TL and provide a framework to guide TL for wide-ranging applications in science and engineering, such as climate change modeling.
\end{abstract}

\section{Introduction}
There are ever-growing efforts focused on using machine learning (ML), particularly the powerfully expressive deep neural networks (NNs), to improve simulations or predictions of nonlinear, multi-scale, high-dimensional systems. For example, in thermo-fluid sciences and in weather/climate modeling, a number of different approaches using NNs have shown significant promise for fully data-driven forecasting, subgrid-scale (SGS) closure modeling, and novel ways of solving partial differential equations (PDEs) \citep{schneider2017earth,brenowitz2018prognostic,rasp2018deep,bolton2019applications, beck2019deep,ham_deep_2019,raissi2019physics,weyn2020improving,brunton_machine_2020,yuval2020stable,kochkov2021machine,novati2021automating,pathak2022fourcastnet}. However, one major challenge facing such efforts is the inability of NNs, and more broadly ML techniques, to {\it generalize} out-of-distribution, i.e., to perform equally well when tested on a dataset whose distribution (or some measure of its statistics) is different from the training set \citep{yosinski_how_2014,nagarajan2020understanding}\footnote{Throughout this paper, we use ``out-of-distribution'' to indicate cases in which the training and testing datasets have different distributions. Furthermore, we use ``out-of-sample'' for accuracy computed using samples from a testing set that is completely independent from the training set, but has the same distribution.}. Some degree of such generalization (i.e., extrapolation) is essential for NNs to be practically useful in many applications. For instance, NN-based SGS closures (i.e., data-driven parameterizations) should work accurately for a range of climates to be useful for global warming projections. If this were not the case, once some parameters (e.g., sea-surface temperature or forcing) change, the data-driven closures may lead to unstable or inaccurate simulations \citep{rasp2018deep,chattopadhyay_datadriven_2020,beucler2021enforcing}. Studies have found a similar challenge arising across thermo-fluid applications~\citep{subel_data-driven_2021,chung2021interpretable,frezat2021physical,taghizadeh2020turbulence,guan_stable_2021}.\\

Transfer learning (TL) provides a  powerful and flexible framework for improving the out-of-distribution generalization of NNs, and has shown success in various ML applications \citep{tan2018survey,zhuang2020comprehensive,yosinski_how_2014}. Consider a NN that is already trained on a large-enough number of training samples ($M_{tr}$) from a {\it base system} and makes predictions with sufficient out-of-sample accuracy. We hereafter refer to this network as a base NN (BNN). The goal of TL is to build a new NN from a BNN that works with similar accuracy for a {\it target system} whose statistical properties could be different from those of the base system. For instance, this could be because of a change in physical properties (e.g., in the context of turbulence, an increase in Reynolds number, \textit{Re}) or in  external forcing (e.g., in the context of climate change, a higher radiative forcing due to increased greenhouse gases). We refer to this network as a TLNN. In TL, a (usually small) number of the layers of the BNN are re-trained, starting from their current weights, with a {\it small} number of re-training samples from the target system (e.g., $M_{tr}/10$ or $M_{tr}/100$ samples). The TL procedure, if properly formulated (as discussed later), can produce a TLNN whose out-of-sample accuracy for the target system is comparable to that of the BNN, despite using only a small amount of re-training data from the target system. \\

In thermo-fluid sciences and weather/climate modeling, a few studies have reported such success with TL for SGS closure modeling and spatio-temproal forecasting~\citep{chattopadhyay_datadriven_2020, inubushi2020transfer, subel_data-driven_2021, guan_stable_2021, guastoni2021convolutional, yousif2021high}. For example, in data-driven closure modeling with a convolutional NN (CNN) for large-eddy simulation (LES) of decaying 2D turbulence, Guan~{\it et al.}~\citep{guan_stable_2021} showed stable and accurate {\it a posteriori} (online) LES using only $M_{tr}/100$ re-training samples from a target system that had a $16\times$ higher $Re$ number. Aside from enabling generalization for one system when parameters change, TL can also be used to effectively blend datasets of different quality and length for training, e.g., a {\it large}, high-fidelity training set from high-resolution simulations and a {\it very small} but {\it higher}-quality re-training set from observations/experiments or much {\it higher}-resolution simulations~\citep{ham_deep_2019,rasp_datadriven_2021, mondal_transfer_2021,chattopadhyay2022long}. Such an application of TL in blending large climate model outputs and small observational datasets has shown promising results in forecasting El~Ni$\mathrm{\tilde{n}}$o–Southern Oscillation and daily weather~\citep{ham_deep_2019,rasp_datadriven_2021, hu2021deep}. Even further, TL has been suggested as a way to improve the training of physics-informed NNs, a novel PDE-solving technique~\citep{karniadakis2021physics,chakraborty2021transfer}. \\



In the TL procedure, there is one critical decision to make: Which layer(s) to re-train? This is an important question, considering that the goal of TL is to find the best-performing TLNN given the constraint imposed by the limited availability of re-training samples from the target system. Finding the best layer(s) to re-train via trial-and-error can become intractable for deep NNs, given that hyperparameter tuning and {\it a priori} (offline) and {\it a posteriori} (online) tests would be needed for each trial (i.e., a combination of re-trained layers). So far, all of the aforementioned studies using TL for turbulence or weather/climate modeling have followed the conventional wisdom from the ML community \citep{yosinski_how_2014,hussain2018study,talo2019application}, which is to re-train the {\it deepest}, i.e., near the output, layers (or have re-trained all layers or most layers ad-hoc). The idea here, mainly developed based on experiments and analyses using static images, is that the shallow layers learn general features of images while the deep layers learn features specific to the images in a given training set \cite{zeiler2014visualizing}. Thus, for effective TL to an out-of-distribution set of images, these deepest layers are the best to re-train \cite{yosinski_how_2014}. Following this idea of re-training the deepest layers has yielded good results in the aforementioned studies on turbulence and weather/climate modeling, e.g., to extrapolate to canonical flows with 10-16 times higher $Re$ numbers. However, given the increasing interest in using TL, its broad applications in these areas, and the need for effective TL in more complex systems, the best practices and the learned physics should be understood and readily accessible. Specifically, the question of the best layer(s) for re-training should be more deeply investigated for the types of data and networks relevant to turbulence and weather/climate modeling applications. Here, we report on such an investigation for the first time. \\

In this paper, we use CNN-based non-local SGS closure modeling for LES of several setups of forced 2D turbulence as the test case. We first demonstrate the power of TL in enabling extrapolation to $100\times$ higher $Re$ numbers, and even more challenging target flows. We further show that here, against the conventional wisdom in the ML literature, the {\it shallowest} layers are the best to re-train. Next, we leverage the fundamentals of turbulence physics and recent theoretical advances in ML to
\begin{enumerate}
    \item Explain what is learned during TL to a different turbulent flow, which is based around changes in the convolution kernels of the BNN after re-training to the TLNN, and these kernels' physical interpretation,
    \item Explain why the shallowest layers, rather than the deepest ones, are the best to re-train in these setups,
    \item Introduce a general framework to guide TL of similar systems based on a number of analysis steps that could be performed before re-training any TLNN.
\end{enumerate}
While we use the SGS modeling of canonical 2D turbulence as the test case, the methods used for (1)-(2) and the framework in (3) can be readily applied to any other TL applications in turbulence or weather/climate modeling. More broadly, this framework can be used for TL applications beyond SGS modeling and for any multi-scale, nonlinear, high-dimensional dynamical systems.

\section{Background}
 The dimensionless governing equations of 2D turbulence in a doubly periodic square domain are:
\begin{subequations}\label{Navier-Stokes}
\begin{eqnarray}
\frac{\partial \omega}{\partial t} + \underbrace{\frac{\partial \psi}{\partial y}\frac{\partial \omega}{\partial x} - \frac{\partial \psi}{\partial x}\frac{\partial \omega}{\partial y}}_{\mathcal{N}(\omega,\psi)}&=&\frac{1}{Re}\nabla^2\omega - \underbrace{m_f\cos{(m_fx)} + n_f\cos{(n_fy)}}_{f(x,y)} -r\omega, \label{eq:NS1}\\
\nabla^2\psi &=& -\omega, \label{eq:NS2}
\end{eqnarray}
\end{subequations}
where $\psi$ is the stream-function, $\omega$ is the vorticity, and $\mathcal{N}(\omega,\psi)$ is the advection term. $r$ is the linear drag coefficient and $f(x,y)$ is a time-independent external forcing at wavenumbers $m_f$ and $n_f$. This system, with different combinations of $f$ and $r$, is a fitting prototype for a variety of large-scale geophysical and environmental flows and has been widely used to test novel techniques including data-driven SGS closures~\citep{maulik2019subgrid,guan_stable_2021,kochkov2021machine,page2021revealing,pawar2022frame,guan2022learning}.\\

For direct numerical simulations (DNS), Eqs.~\eqref{eq:NS1}-\eqref{eq:NS2} are solved using a pseudo-spectral solver with high resolution ($N_{\rm{DNS}}$ collocation grid points in each direction), resolving all relevant spatio-temporal scales (see Section~\ref{numerical simulation} for the solver's details). Filtering Eqs.~\eqref{eq:NS1}-\eqref{eq:NS2} yields equations for LES (Eqs.~\eqref{eq:FNS1}-\eqref{eq:FNS2}). In the LES equations, an SGS term, $\Pi =  \mathcal{N}(\overline{\omega},\overline{\psi}) - \overline{\mathcal{N}({\omega},{\psi})}$, arises and has to be explicitly represented in terms of the resolved flow $(\overline{\omega},\overline{\psi})$ via an SGS closure. Here, $\overline{(\cdot)}$ denotes filtering and coarse-graining (see Section~\ref{sec:filtering} for details). The same pseudo-spectral solver, but with lower spatio-temporal resolution (e.g., $N_{\rm{LES}} = N_{\rm{DNS}}/8$ and a $10\times$ larger time step), is used to solve the LES equations~(\eqref{eq:FNS1}-\eqref{eq:FNS2}). While the LES solver is computationally much cheaper, it requires an accurate closure for $\Pi(\overline{\omega},\overline{\psi})$, a long-standing challenge in every discipline of science and engineering dealing with turbulent flows.  \\

Here, to build data-driven closures, we train CNNs on filtered and coarse-grained DNS (FDNS) data\footnote{Whether the training FDNS data are from the base or target system or both is clearly explained for each analysis.}: The input of the CNNs is $(\bar{\psi},\bar{\omega})$ and the output is $\Pi$ (see Sections~\ref{sec:filtering} and \ref{sec:CNN} for details). By changing $Re$, $r$, $m_f$, and $n_f$, we have created 6 distinctly different flows, divided into 3 cases, each with a base and a target system (Table~\ref{tab:table_setups} and Section~\ref{sec:Systems}). We have shown in previous studies that for various setups of 2D turbulence, CNNs trained on large training sets, or on small training sets with physics-constraints incorporated, produce accurate and stable data-driven closures in {\it a priori} (offline) and {\it a posteriori} (online) tests~\citep{guan_stable_2021,guan2022learning}. These CNN-based closures were found to accurately capture both diffusion and backscattering, and to outperform widely used physics-based SGS closures such as the Smagorinsky, dynamic Smagorinsky, and mixed models in both {\it a priori} and {\it a posteriori} tests. In this paper, we focus on TL and addressing objectives (1)-(3) listed in the Introduction.

\section{Results}
\label{sec:Results}
\subsection{The ability of TL in closing large generalization gaps in SGS modeling}
\label{sec:Online_Results}
\begin{figure}
    \centering
    \includegraphics[width=\textwidth]{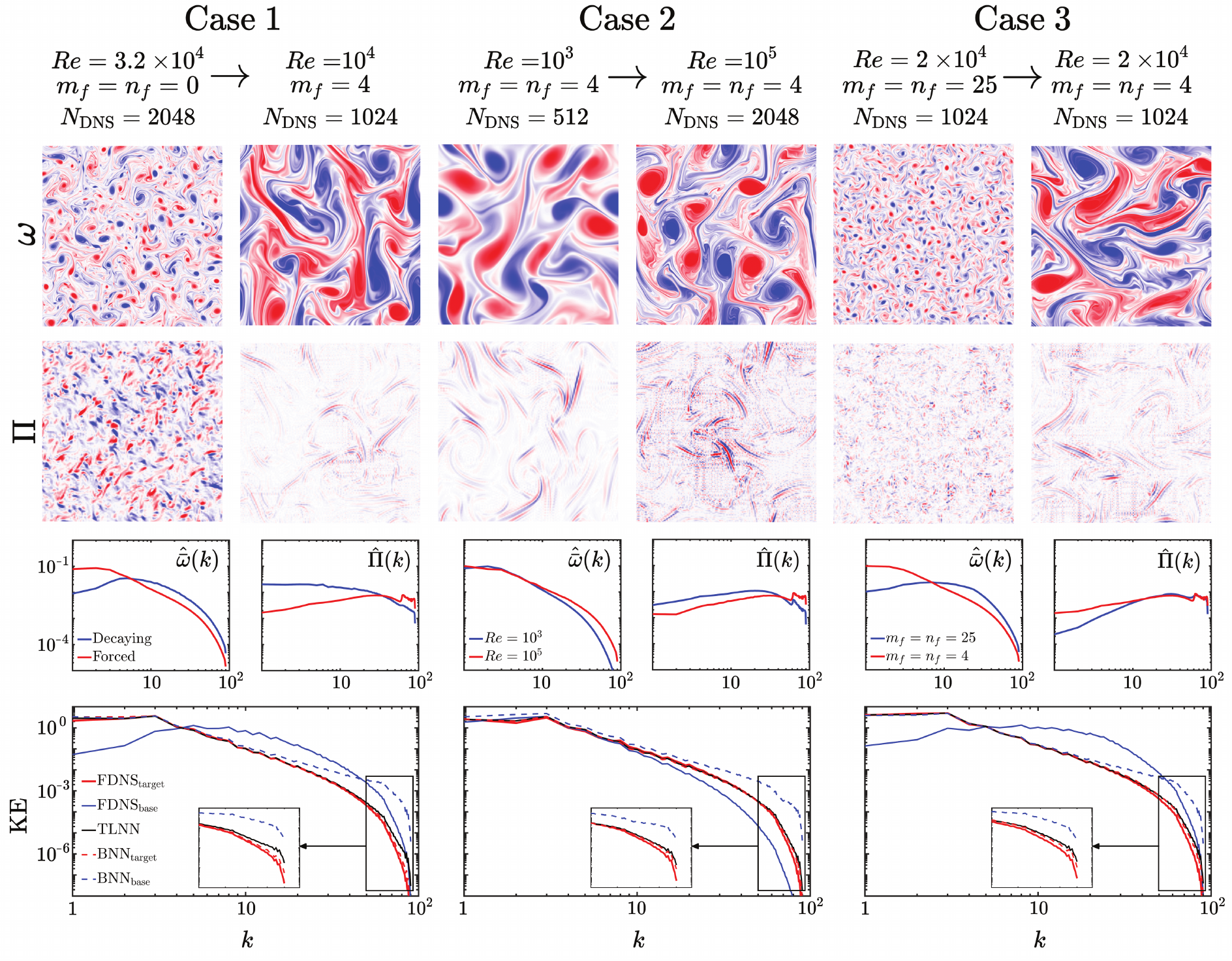}
    \caption{Some comparisons between the base and target systems of the three cases (rows~1-3) and the ability of TL to close the generalization gaps in {\it a posteriori} (online) LES (row~4). Parameters of the 6 systems are listed in Table~\ref{tab:table_setups}, and these cases are described in Section~\ref{sec:Systems}. Each case consists of a base (left column) and a target (right column) system. The first and second rows show, respectively, the DNS snapshots of one of the inputs to the CNNs, $\omega$, and the snapshots of the SGS terms, $\Pi$, the output of the CNNs (note that $N_{\rm{LES}}=128$ for all systems). These rows visualize the substantial differences in the length scales dominating the base and target systems in each case. To further demonstrate these differences in spatial scales, using the entire training sets and solid blue lines for base and solid red lines for target systems, we show the angle-averaged spectra of $\bar{\omega}$ (left) and $\Pi$ (right) in the third row, and the KE spectra of FDNS in the fourth row. In these panels, the horizontal axis is wavenumber $k=\sqrt{k_x^2+k_y^2}$, where $k_x$ and $k_y$ are the wavenumbers in $x$ and $y$ directions. The fourth row also shows the {\it out-of-sample} accuracy of the NN-based closures: The KE spectra are from {\it a posteriori} LES of the {\it target} systems using SGS closures that are BNNs trained on $M_{tr}$ samples from the {\it base} systems (BNN$_{base}$, dashed blue lines) or from the {\it target} systems (BNN$_{target}$, dashed red lines), or from the TLNN (black lines) re-trained using $M_{tr}/10$ samples~(see Section~\ref{sec:CNN} for details). In all three cases, there is a large generalization gap (difference between the dashed blue and solid red lines), particularly for $k>10$. In each case, TL closes this gap (black and solid red lines almost overlap for all $k$). Note that for the TL here, layers 2 and 5 are re-trained for Case~1, and layer~2 is re-trained for Cases 2 and 3 (see Section~\ref{sec:isolating} and Fig.~\ref{fig:Emperical_Results} for more discussions). }
    \label{fig:figure_1}
\end{figure}

Before attempting to explain the physics of TL, we first show that TL enables our CNN-based SGS closures to effectively generalize between the base and target systems in each of the three cases. The first three rows of Fig.~\ref{fig:figure_1} demonstrate the differences in spatial scales between each pair of base and target systems. In Case~1, the base system is decaying turbulence while the target systems is forced turbulence. From the ${\omega}$ and $\Pi$ snapshots, their spectra, and the kinetic energy (KE) spectra, it is clear that the two systems are different at both the large and small scales. As a results of these substantial differences across all scales, the LES of the target system using a BNN trained on the base system (BNN$_{base}$) produces a KE spectrum that does not agree with that of the target system's FDNS (the truth). This indicates that the BNN$_{base}$ fails to generalize here, leading to a generalization gap that is the difference between the two KE spectra (most noticeable at wavenumbers, $k$, larger than 10). Note that comparing the KE spectra of FDNS and LES is the most common measure of the {\it a posteriori} (online) performance of SGS closures. \\

Similar failures of the BNN$_{base}$ to generalize are seen for Cases~2 and 3, leading to large generalization gaps in the KE spectra. In Case~2, the base system has $Re=10^3$ and the target system has $Re = 10^5$. This $100\times$ increase in the $Re$ number leads to the development of more small-scale features in the target system, and changes the spectrum of $\Pi$ in both large and small scales. In Case~3, the forcing of the base system is at wavenumber $m_f=n_f=25$, while the target system's forcing is at $m_f=n_f=4$. This decrease in forcing wavenumbers results in more (less) large-scale (small-scale) structures in the resolved flow, as seen in the spectra of both $\bar{\omega}$ and KE. This change in forcing wavenumber also leads to more large-scale structures in $\Pi$ without any noticeable change in its small-scale structures. In short, Cases 1-3 represent 6 fluid flow systems that are different in terms of both the physics that drive the differences and the spatial scales of the resolved and SGS components. \\

In all three cases, TL closes the {\it out-of-distribution} generalization gap: LES of the target system using a TLNN (re-trained with $M_{tr}/10$ samples) produces a KE spectrum that matches that of the target system's FDNS. For the LES of the target system, the TLNN not only significantly outperforms the BNN$_{base}$, but is almost as good as the BNN trained on $M_{tr}$ samples from the target system, BNN$_{target}$ (see the insets in Fig.~\ref{fig:figure_1}).



\subsection{Impact of the choice of the re-trained layer on offline and online accuracy}\label{sec:isolating}
Figure~\ref{fig:figure_1} shows the power of TL in closing the generalization gaps. These results also show that in contrast to conventional wisdom, the best layers to re-train are not the deepest, but rather, the {\it shallowest} ones. For each case, we have explored all 1-, 2-, and 3-layer combinations for re-training. Based on the correlation coefficient of the $\Pi$ terms from FDNS and TLNN, which is the most common metric for {\it a priori} (offline) tests, we have found that for Cases 2 and 3, re-training layer~2 alone is enough to get the best performance. For Case~1, re-training layers 2 and 5 provides the best performance, although most of the gap can be closed by re-training layer 2 alone.   \\

\begin{figure}
    \centering
    \includegraphics[width=\textwidth]{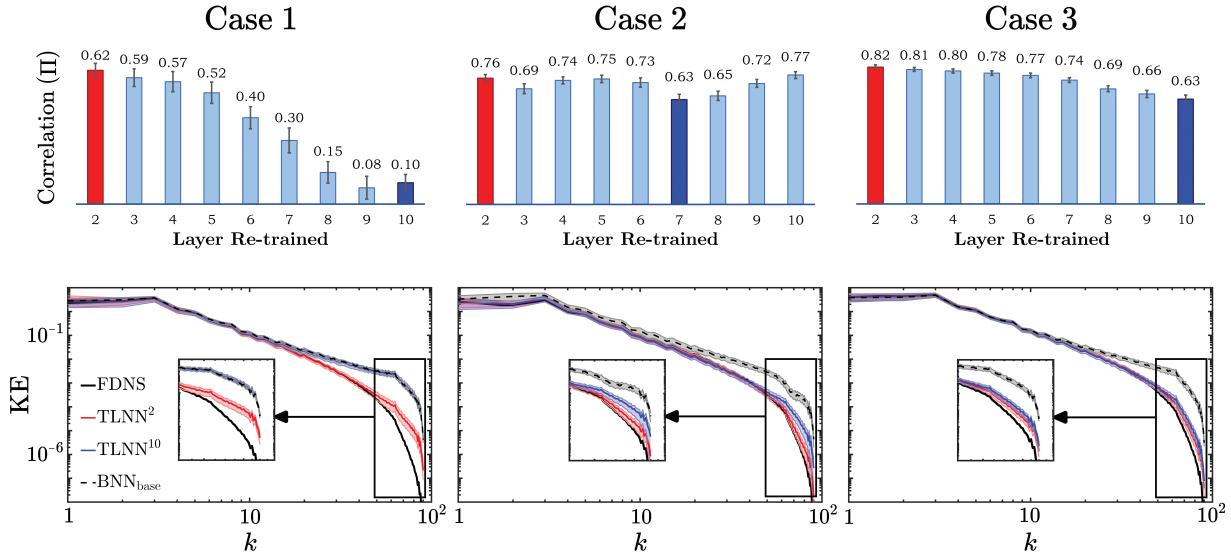}
    \caption{Online and offline performance of TLNNs as a function of the individual re-trained layer. For each individual layer re-trained with $M_{tr}/10$ samples, the top row shows the most common measure of {\it a priori} (offline) accuracy of a SGS model: the correlation coefficient between $\Pi$ from FDNS (truth) and from the TLNN. The vertical lines on the bar plots show uncertainty measured as the standard deviation calculated over $100$ random samples from the testing set. The bottom row shows the KE spectra of the target systems' FDNS and the KE spectra from {\it a posteriori} (online) LES with BNN$_{base}$ or TLNN$^\ell$, where $\ell$ indicates the re-trained layer. These KE spectra are calculated using $5$ long integrations, each equivalent to $10^6 \Delta t_{\mathrm{DNS}}$. Shading shows uncertainty, estimated as 25$^{th}$-75$^{th}$ percentiles of standard error calculated from partitioning each of the 5 runs into $10$ sub-intervals. For each case, the best (worst) individual layer to re-train is shown in red (blue) in both rows. The best and worst performing layers here are chosen based on the online performance, i.e., how closely the KE spectrum matches that of the FDNS. Note that in Fig.~\ref{fig:figure_1}, both layers 2 and 5 are re-trained during TL for Case~1, leading to a better TLNN with LES' KE spectrum matching that of the FDNS even at the highest wavenumbers.}
    \label{fig:Emperical_Results}
\end{figure}

 To better understand the effects of ``re-training layer'' selection in TL, Fig.~\ref{fig:Emperical_Results} shows the offline and online performance of TLNN$^\ell$ as a function of an individual re-trained hidden layer $\ell$. In Case~1, the offline performance of TLNNs substantially declines as deeper layers are used for re-training (top row). As a result, TL with deepest layers is completely ineffective; for example, LES with TLNN$^{10}$ is as poor as LES with BNN$_{base}$, leaving a large generalization gap in the KE spectrum for $k>10$ (bottom row). In contrast, LES with TLNN$^2$ has a KE spectrum that closely matches that of the FDNS and only has a small generalization gap for $k>40$ (as shown in Fig.~\ref{fig:figure_1}, this gap is further closed when both layers 2 and 5 are re-trained). Similarly, in Case~3, the offline performance of TLNNs declines as $\ell$ increases. That said, in this case, TL with even the worst layer to re-train ($\ell=10$) is effective in closing the generalization gap in the online test. Still, LES with TLNN$^2$ is slightly better than LES with TLNN$^{10}$ (see the inset). In these two cases, there are substantial changes in the large scales of the inputs and outputs between the base and target systems (see the spectra of $\bar{\omega}$ and $\Pi$ in Fig.~\ref{fig:figure_1}). The offline results show a clear deterioration of the performance when moving from shallow to deep layers, which is due to the inability of the deeper layers to learn about changes in large scales during TL, as shown later. \\

 In Case~2, the offline performance of TL is not a monotonic function of $\ell$, though $\ell=2$ is still the best layer to re-train ($\ell=7$ is the worst), based on both offline and online results. The non-monotonicity emerges because changes between the base and target systems' $\bar{\omega}$ and $\Pi$ occur predominantly at smaller scales (see their spectra in Fig.~\ref{fig:figure_1}), which deeper layers are also able to learn during TL. For this case, as in Case~1, there is a noticeable difference in the online performance of the LES with TLNNs that use the best and worst performing re-trained layers.   \\




The above analysis demonstrates that a poor selection of the re-training layer can lead to poor offline and/or online performance of the TLNN. This analysis also shows that in all three cases, re-training the shallowest layers consistently yields the best-performing TLNNs. This is in contrast to the conventional wisdom of TL, which is predominantly built on studies on static images that often do not have a continuous spectrum of  spatial scales~\cite{neyshabur_what_2021,yosinski_how_2014,zhuang_comprehensive_2021}.

\subsection{Failure of deep layers to learn changes in large scales during TL}
\label{sec:activations}
\begin{figure}
    \centering
    \includegraphics[width=\textwidth]{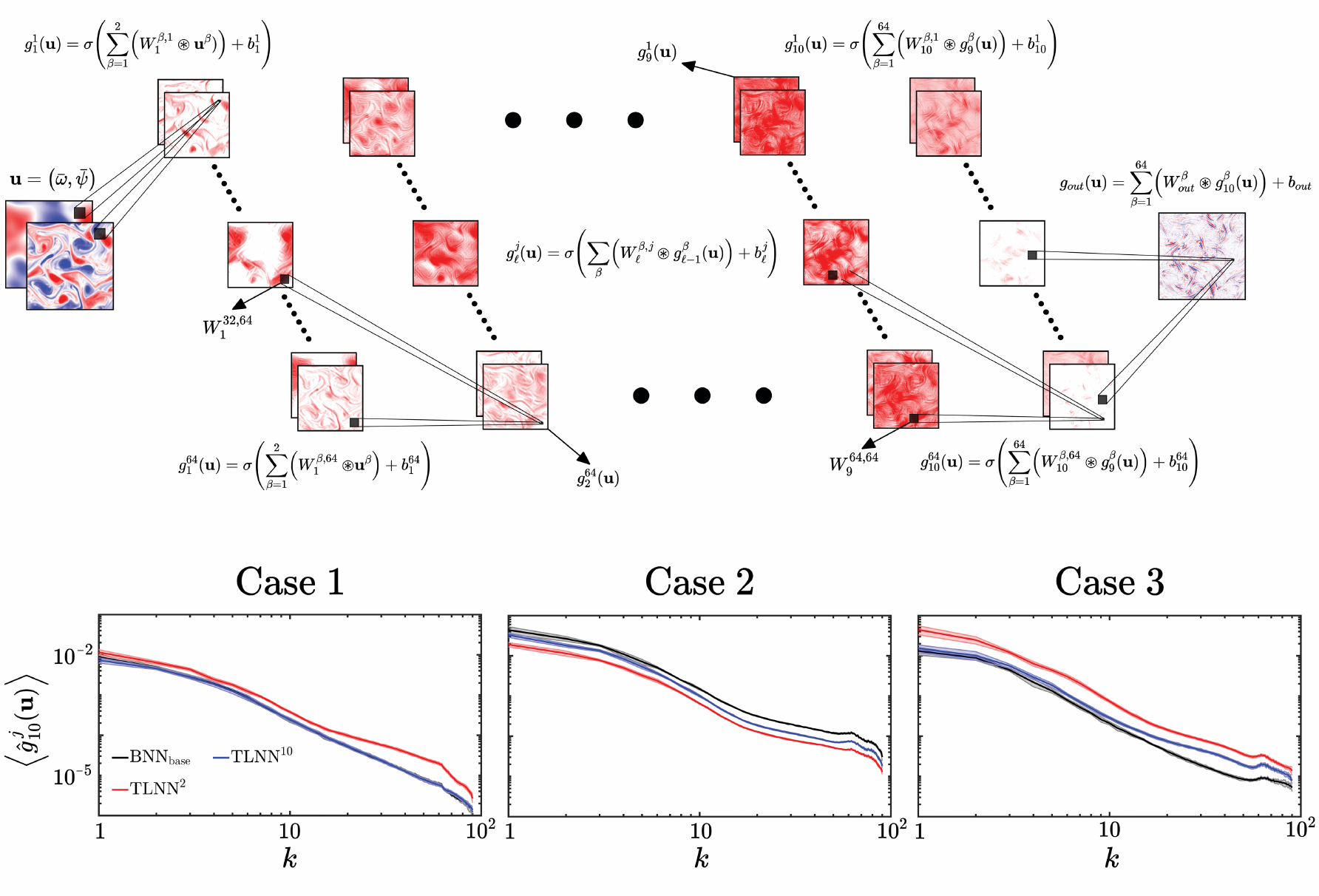}
    \caption{The top row shows a schematic of the CNN architecture and its governing equations. Examples of activations $g^j_\ell \in \mathbb{R}^{{128} \times {128}}$ of some of the layers $\ell$ and channels $j$ are shown as red shading (with $\sigma$ being the ReLU nonlinear function, the values of these activations are all positive). Note that training a CNN means learning the convolution kernels' weight matrices $W^{\beta,j}_\ell \in \mathbb{R}^{{5} \times {5}}$ and biases' constant matrices $b^j_\ell \in \mathbb{R}^{128 \times 128}$,  (for hidden layers $\ell=2 \cdots 10$, $\beta \in \{1, 2 \cdots 64\}$ and $j \in \{1, 2 \cdots 64\}$). See Section~\ref{sec:CNN} for a detailed discussion of the CNN and its mathematical representation. In the bottom row, the effects of re-training layer 2 versus layer 10 on the Fourier spectrum of the averaged activation of the last hidden layer ($\ell=10$) are compared (note that the output layer $\ell=11$ has a linear activation function). The averaging is done over all channels, denoted by $\left< \cdot \right>$. Shading shows uncertainty, estimated as 25$^{th}$-75$^{th}$ percentiles of the averaged activation spectra computed with 20 random input samples.} 
    \label{fig:activations}
\end{figure}

To understand why different re-training layers lead to different TL performance, next, we conduct a spectral analysis of the CNNs in this section and the next one. The mathematical representation of CNNs is discussed in Section~\ref{sec:CNN}. Explained briefly, in our CNNs, inputs $\mathbf{u} = \left(\bar{\omega}, \bar{\psi} \right)$ are passed through 11 sequential convolutional layers to predict outputs, $\Pi$ (Fig.~\ref{fig:activations}). The hidden layers each have 64 channels. The output of channel $j$ of layer $\ell$, called activation $g^j_\ell$, is computed using Eq.~\eqref{eq:activations}: 64 kernels perform convolution on  $g^j_{\ell-1}$ of each of the 64 channels, $j$, and the outcome of these linear operations is sent through a ReLU nonlinear activation function, $\sigma$. Figure~\ref{fig:activations} shows examples of $g^j_\ell$, which are $128 \times 128$ matrices. Note that these $64^2$ kernels in each hidden layer extract information from the activations through spatial convolution, and their weight matrices $W^{\beta,j}_\ell \in \mathbb{R}^{5\times 5}$ are the main parameters that are learned during the training of a CNN. \\

In the second row of Fig.~\ref{fig:activations}, we compare the all-channels-averaged Fourier spectra of activations of the last hidden layer $\left<\hat{g}^j_{10}\right>$ from a fully trained BNN$_{base}$, TLNN$^2$, and TLNN$^{10}$ ($\left< \cdot \right>$ represents averaging over all channels and $\hat{\cdot}$ means Fourier transform). The spectrum of $\left<\hat{g}^j_{10}\right>$ from TLNN$^{2}$ differs from that of the BNN$_{base}$ at most wavenumbers including the small wavenumbers. This indicates that re-training layer 2 {\it can} account for differences in the output ($\Pi$) from the base and target flows at all scales, including the large scales. In contrast, the spectra from TLNN$^{10}$ are almost the same as those from BNN$_{base}$ at all scales (Case 1) or at large scales $k<10$ (Cases 2 and 3). This indicates that re-training layer 10 {\it cannot} account for differences in the output from the base and target flows at large scales. Given that in all 3 cases there are large-scale differences in the $\Pi$ terms between the base and target flows (Fig.~\ref{fig:figure_1}), this analysis explains why re-training layer 10 (or other deep layers) leads to ineffective TL, while re-training layer 2 leads to the best TL performance.  \\

To further understand what controls the spectra of $g^j_\ell$, we have examined Eq.~\eqref{eq:gspectra}, which is the analytically derived Fourier transform of Eq.~\eqref{eq:activations}. As discussed in Section~\ref{sec:spec_analysis}, this analysis shows that the Fourier spectrum of ${g}^j_\ell$ depends on the spectrum of $\hat{g}^j_{\ell-1} \in \mathbb{C}^{128\times 128}$, the spectra of the weight matrices $\hat{\widetilde{W}}_\ell^{\beta,j} \in \mathbb{C}^{128\times 128}$ (and constant biases $\hat{b}_\ell^j \in \mathbb{R}$), as well as where linear activation $h^j_{\ell}(x,y)>0$ (defined in Eq.~\eqref{eq:act_linear}). The latter is a result of the Fourier transform of the ReLU activation function, the only source of nonlinearity in the calculation of $g^j_\ell$. In Supplementary Figure~1, we have compared the spectra of activations from layers 2 and 10 before and after applying the ReLU activation function. From this, we find that in all 3 cases, linear changes due to updating the weights substantially alter the spectra of the activations while nonlinear changes only play a significant role in Case~1. These results (and further discussions in Section~\ref{sec:spec_analysis}) suggest that a deeper insight into TL might be obtained by examining the spectra of the weight matrices, $\hat{\widetilde{W}}_\ell^{\beta,j}$, and how they change from BNN$_{base}$ to TLNN, as done in the next section.

\subsection{Spectral analysis of kernels' weights}
\label{sec:specanalysis}

Before investigating how TL changes the spectra of kernels' weights, let's first look at the spectra from the BNN$_{base}$ of the 3 cases. A close examination of $\left|\hat{\widetilde{W}}_\ell^{\beta,j}\right|$ in different layers shows that the learned kernels are a combination of low-, band-, and high-pass filters. While visualizing all the $64^2$ kernels in each layer is futile, we realize that the similarity across the spectra of many kernels allows us to meaningfully cluster them using the $k$-means algorithm. Supplementary Figure~2 presents the cluster centers (in Fourier space) for $\ell=2$ and $10$ for each case. \\


Since deep CNNs contain a very large number of parameters ($O(10^6)$), it is often intractable to isolate the effect of each convolution kernel for either a BNN or TLNN. Moreover, investigating the learned convolution kernels in physical space ($W^{\beta,j}_\ell \in \mathbb{R}^{5 \times 5}$) does not lead to any meaningful physical understanding. Above, we show that examining the kernels in the spectral space ($\hat{\widetilde{W}}^{\beta,j}_\ell \in \mathbb{C}^{128 \times 128}$) leads to physically interpretable insight into their role as spectral filters. Still, due to the large number of parameters and the impact of nonlinearities, it is currently challenging to understand the physics learned by the entire BNN. Fortunately, due to the over-parameterized nature of these deep CNNs, TL occurs in the \textit{lazy training regime} \cite{chizat2019lazy}. In this regime, significant changes occur in only a small number of kernels, as shown below. This opens an avenue for {\it explaining what is learned in TL} through examining the spectra of the few kernels with the largest changes. \\

\begin{figure}
    \centering
    \includegraphics[width=\textwidth]{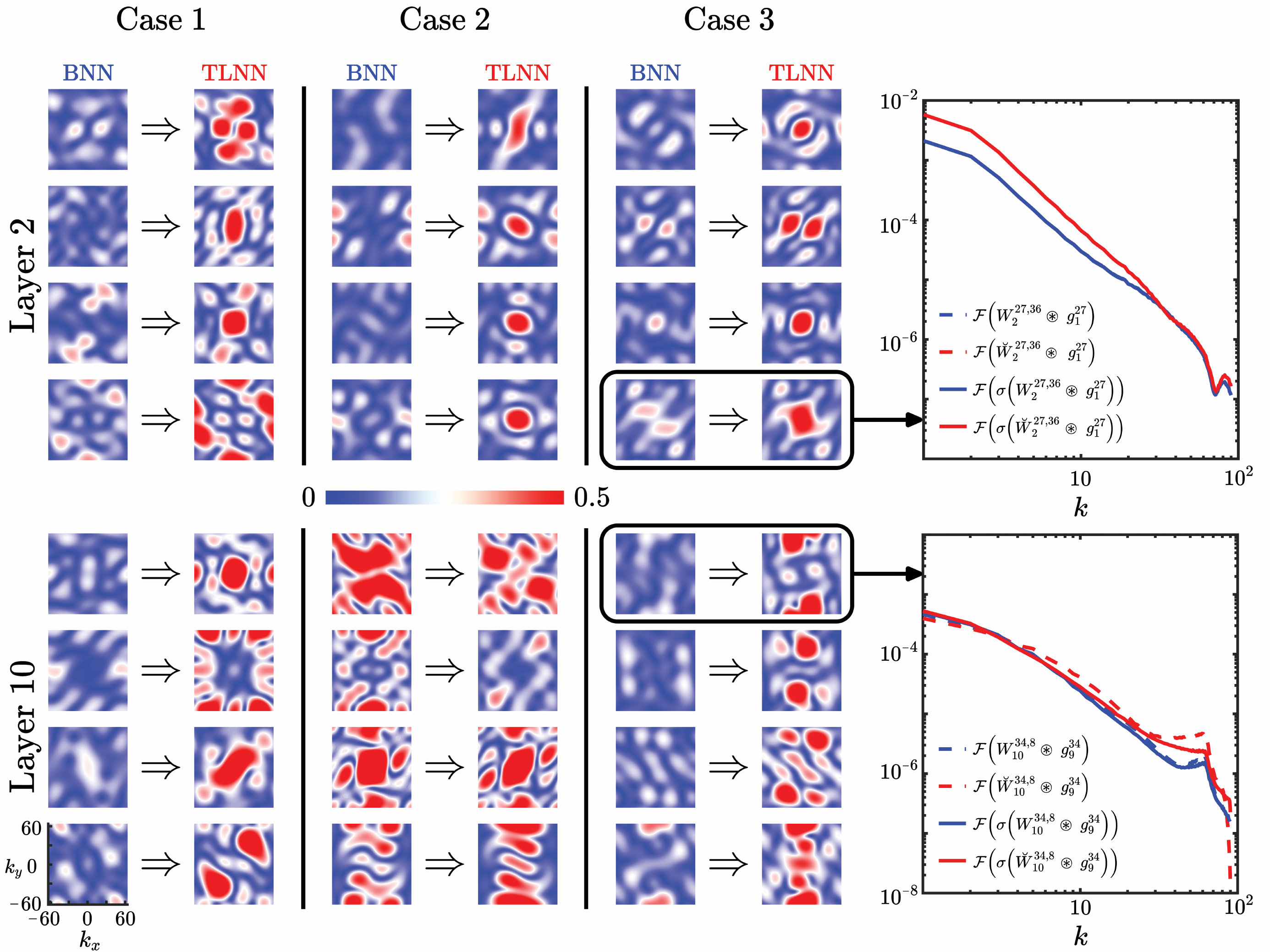}
    \caption{The 3 left columns compare the Fourier spectra $\left|\widetilde{W}_\ell^{\beta,j}\right|$ of the 4 convolution kernels that have changed the most between BNN$_{base}$ and TLNN$^2$ (top row) and TLNN$^{10}$ (bottom row). The change in each kernel is quantified using the Frobenius norm $\left \Vert \mathcal{F}\left(\breve{\widetilde{W}}_\ell^{\beta,j}\right)-\mathcal{F}\left({\widetilde{W}}_\ell^{\beta,j}\right) \right \Vert_F$, where $\mathcal{F}$ indicates the Fourier transform (Eq.~\eqref{eq:fft_normal}) and $\breve{\cdot}$ indicates that the weight matrix is from a TLNN (absence of $\breve{\cdot}$ in this figure means that the matrix is from a BNN$_{base}$). The two panels on the right show examples of how changes in one kernel of layer 2 and one of layer 10 affect the activations' spectra of layer~10 by comparing $\hat{g}^j_{10}$ from BNN$_{base}$ (solid blue) with that from the TLNN$^\ell$ (solid red). We also show the activations before the application of ReLU nonlinearity $\sigma$ with dashed lines. Note that the inputs to the networks ($\mathbf{u}$) are the same and from the {\it target} system. The top panel shows that the newly learned kernel in layer~2 substantially changes the activation in low wavenumbers $(k \le 20)$ without affecting the higher wavenumbers, as expected from a low-pass filter. Here, nonlinearity has little impact: the solid and dashed lines coincide. The bottom panel shows that the newly learned kernel in layer~10 only changes the activation at high wavenumbers and that in this case, the ReLU nonlinearity has a contribution.}
    \label{fig:Spectra_Comp}
\end{figure}

For each case, we quantify the change in each kernel by computing the Frobenius norm of the difference between $\hat{\widetilde{W}}^{\beta,j}_\ell$ from the BNN$_{base}$ and TLNN$^\ell$ for $\ell=2$ and $10$. As demonstrated in the Supplementary Figure~3, in each case and each layer, there are a few kernels with substantial changes, much larger than the changes in the rest of the $64^2$ kernels. Figure~\ref{fig:Spectra_Comp} shows the spectra of the 4 most-changed kernels (due to TL) in layers 2 and 10 from BNN$_{base}$ and TLNN$^\ell$. We see that in all 3 cases, re-training layer~2 converts a few relatively inactive kernels into clear low-pass filters (one exception is the 4th most-changed kernel in Case~1, discussed later). In contrast, re-training layer~10 turns inactive or complex filters into other complex (often less coherent) filters, though some of them can be identified as band- or high-pass filters. The two panels on the right further show that the kernels learned in TL act as their spectra suggest: the new low-pass filter learned from re-training layer~2 produces activation $g^j_2$ that is different from that of the BNN$_{base}$ (for the same input $\mathbf{u}$) only in large scales, while the most-changed kernel from re-training layer~10 (a high-pass filter) produces activation $g^j_{10}$ that is different from that of the BNN$_{base}$ mainly in the small scales. \\

We remind the reader of the discussion in Section~\ref{sec:activations}: TL needs to capture changes in large scales of the output $\Pi$ between the base and target systems, and the inability of the re-trained layer~10 to do so is the reason for the ineffectiveness of TLNN$^{10}$. Based on the above analysis, we can now explain the reason of this ineffectiveness (and the effectiveness of layer~2): layer~10 fails to learn new low-pass filters, which are essential for capturing changes in the large scales, especially at the end of the network right before the {\it linear} output layer. In contrast, layer~2 is capable of learning new low-pass filters to capture these changes in the large scales of the base and target systems' outputs. Admittedly, the nonlinearity and subsequent layers after $\ell=2$ could impact the outcome of a low-pass filter, but it is possible to separate out the impact of the nonlinearity. Figure~\ref{fig:Spectra_Comp} and Supplementary Figure~1 show the impact of the ReLU nonlinearity by comparing the spectrum of the activation before and after ReLU is applied. In Case 1, where the ReLU function plays an important role in changing the activations' spectra after TL, we find that in addition to low-pass filters, TLNN$^2$ also learns more complex filters, such as the 4th most-changed kernel in Fig.~\ref{fig:Spectra_Comp}, that impact the sign of the linear activations, $h^j_2$. \\

The analyses presented so far provide answers to objectives~1-2 from the Introduction. To address objective~3 (develop a general framework to guide TL), we need to understand why layer~10 cannot learn the filters needed for the TL in these cases while layer~2 can. This question is investigated next by leveraging recently developed ideas in theoretical ML.

\subsection{Loss landscapes: Target system's data and perturbed re-training layer(s)}
\label{sec:landscape}
So far, we have presented \textit{post-hoc} analyses, investigating changes in the spectra of activations and weights, as well as the learned physics, after a BNN$_{base}$ has been re-trained to obtain a TLNN. Here, we present a \textit{non-intrusive} method for gaining insight into which layers of a BNN$_{base}$ are the best (or worst) to re-train for a given target system before performing any actual re-training. This analysis exploits the concept of ``loss landscapes'' \cite{li_visualizing_2018,neyshabur_what_2021,krishnapriyan2021characterizing} and examines, for a given CNN input $\mathbf{u}$, the sensitivity of the loss function $\mathcal{L}$ to perturbations of the weights (and biases) of the layer(s) to be re-trained. Training a deep CNN requires solving a high-dimensional non-convex optimization problem, for which the smoothness of the loss function can be a significant factor in the success of training. Previous studies \cite{li_visualizing_2018,neyshabur_what_2021,krishnapriyan2021characterizing} show that even one-- or two-dimensional approximations of the loss landscape can provide meaningful information about how easily a CNN can be trained. In this study, leveraging recent work in theoretical ML \cite{neyshabur_what_2021}, we extend the application of loss landscape analysis to studying TL; see Section~\ref{sec:loss_land} for details and discussions.   \\

\begin{figure}
    \centering
    \includegraphics[width=\textwidth]{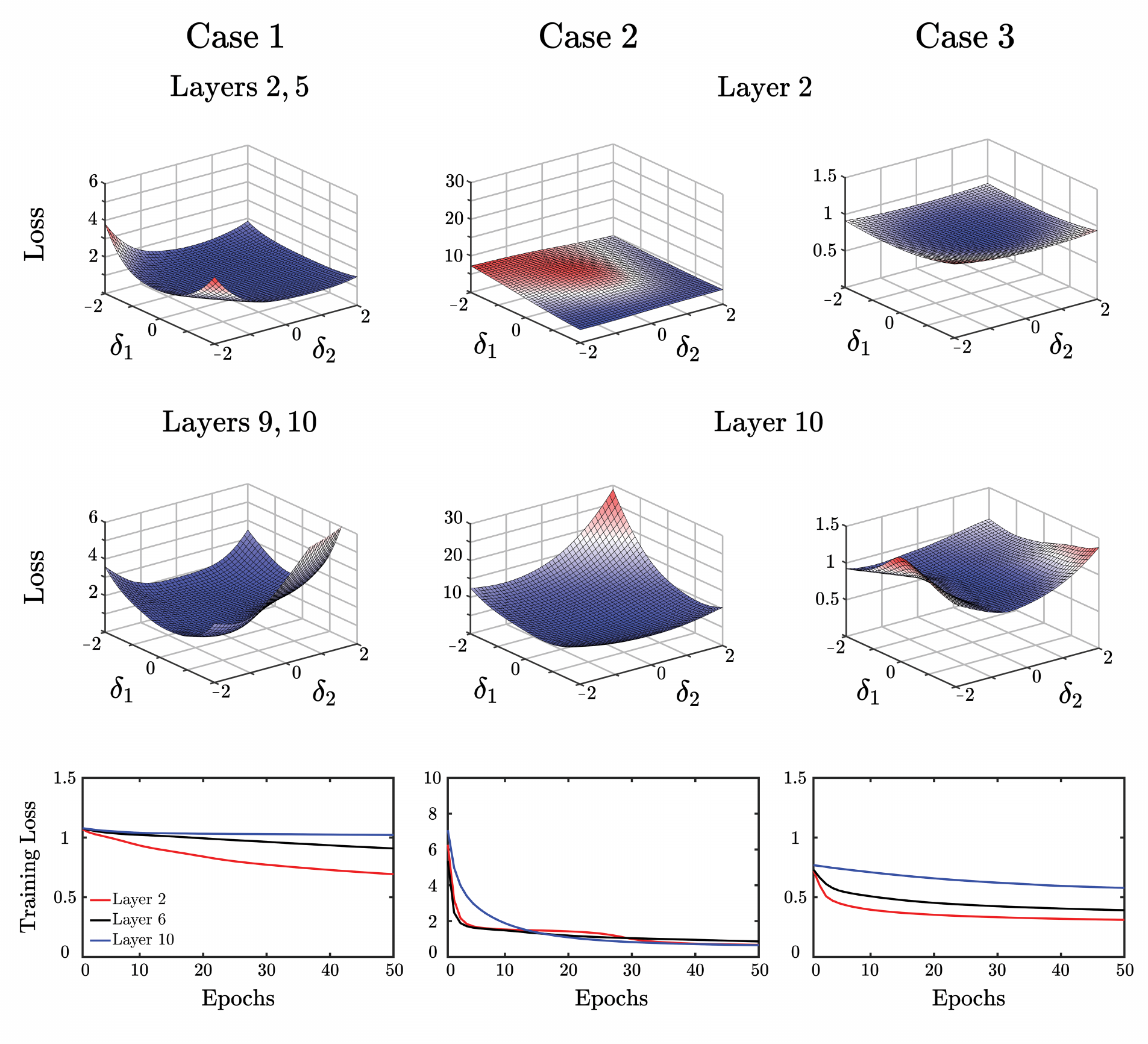}
    \caption{The top two rows present the loss landscape $\mathcal{L}_{(\delta_1,\delta_2)}$ computed from Eq.~\eqref{eq:landscape}. In row~1, the weights and biases of layers $2$ and $5$ (Case 1) or $2$ (Cases~2 and 3) from the BNN$_{base}$ are perturbed in two random directions by amplitudes $\delta_1$ and $\delta_2$; see Section~\ref{sec:loss_land} for details. Similarly, in row~2, the deepest layers are perturbed. Row~3 shows the convergence of the training loss when individual shallow, middle, and deep layers are re-trained for TL. In all calculations, the inputs are from the {\it target} system.}
    \label{fig:Loss}
\end{figure}

Figure~\ref{fig:Loss} (rows~1 and 2) shows the loss landscape calculated for perturbations along two random directions in parameter space of shallow or deep layers for the BNN$_{base}$ with data from the target system as the input. Supplementary Figure 4 presents the loss landscapes obtain using a second method (based on perturbations along the eigenvectors of the Hessian of the loss). These loss landscapes provide insight to indicate if a layer is receptive to change when re-trained with new data during TL. Two important characteristics of these landscapes are their convexity and the magnitude. Notably, the landscapes in row~1 (re-training layer 2, or 2 and 5) are both smooth and of much lower magnitude than those in row~2 (deep layers). For case 1, we show results for combinations of two layers as this yields better performance than re-training a single layer, and this also demonstrates that the method is robust beyond perturbations of individual layers. This analysis indicates that these shallow BNN$_{base}$ layers are easier to re-train for these target systems' data, and that the loss function will likely reach a better optimum during TL. This loss landscape analysis is consistent with our previous findings of TLNN$^2$'s ability (TLNN$^{10}$'s inability) to perform well in these TL tasks. \\

 Additionally, Fig.~\ref{fig:Loss} (bottom row) shows how quickly the loss function decreases as a function of the number of epochs during re-training layer 2, 6, or 10 of the BNN$_{base}$ using the target system's data. For all 3 cases, TLNN$^2$ converges the fastest. This is a direct consequence of the structure of the loss landscapes shown in rows~1 and 2 of Fig.~\ref{fig:Loss}: landscapes obtained from perturbing layer 2 are more favorable for convergence (an absence of pathological non-convexities) as compared to the landscapes obtained from perturbing layer~10. \\

 As a final note, we point out that the concept of ``spectral bias`` \cite{rahaman2019spectral} from theoretical ML suggests that layer~2, which converges faster, is learning the large scales while the slow-converging layer~10 is learning the small scales. This is consistent with the conclusions of our earlier analyses of the weights' spectra in Section~\ref{sec:specanalysis}.



\section{Discussion}
In Section~\ref{sec:Results}, we present a number of novel analysis steps, ranging from a) the most intrusive, computationally expensive ones to gain insight into the learned physics, to b) non-intrusive, inexpensive analysis, which can effectively guide TL for any new problem. For (a), we examine the BNNs' and TLNNs' activations and weights (done after re-training), revealing that the newly learned kernels are meaningful spectral filters, consistent with the physics of the base and target systems and their difference in the spectral space. To the best of our knowledge, this is the first full interpretation of CNNs' kernels in an application for turbulence or weather/climate modeling. For (b), we introduce a novel use of loss landscapes, shedding light on which layers are most receptive to learn the new filters in re-training. \\

These steps connect the spectral analysis of turbulent flows\footnote{Spectral analysis has been the cornerstone of understanding turbulence physics since the pioneering work of Kolmogorov~\cite{kolmogorov1941local}.} and CNNs, and further connect them to the most recent advances in analyzing deep NNs. The above analyses show that the shallowest layers are the best to re-train here, and shed light on the learned physics and the inner-workings of TL for these 3 test cases. Admittedly, some or all of these findings, in terms of learned physics and best layer(s) to re-train, are likely specific to these 3 cases, our specific NN architecture, and the SGS modeling application. However, the analysis methods we introduce or employ are all general and can be used for any base-target systems, applications (SGS modeling, data-driven forecasting, or blending training sets), and most CNN architectures\footnote{The weights' spectra analysis might have to be further modified for networks that involve max-pooling.}. Therefore, putting all these analysis steps together, below we propose a general framework for guiding and explaining TL, which we expect to benefit a broad range of applications involving multi-scale nonlinear dynamical systems. \\

\begin{figure}
    \centering
    \includegraphics[width=\textwidth]{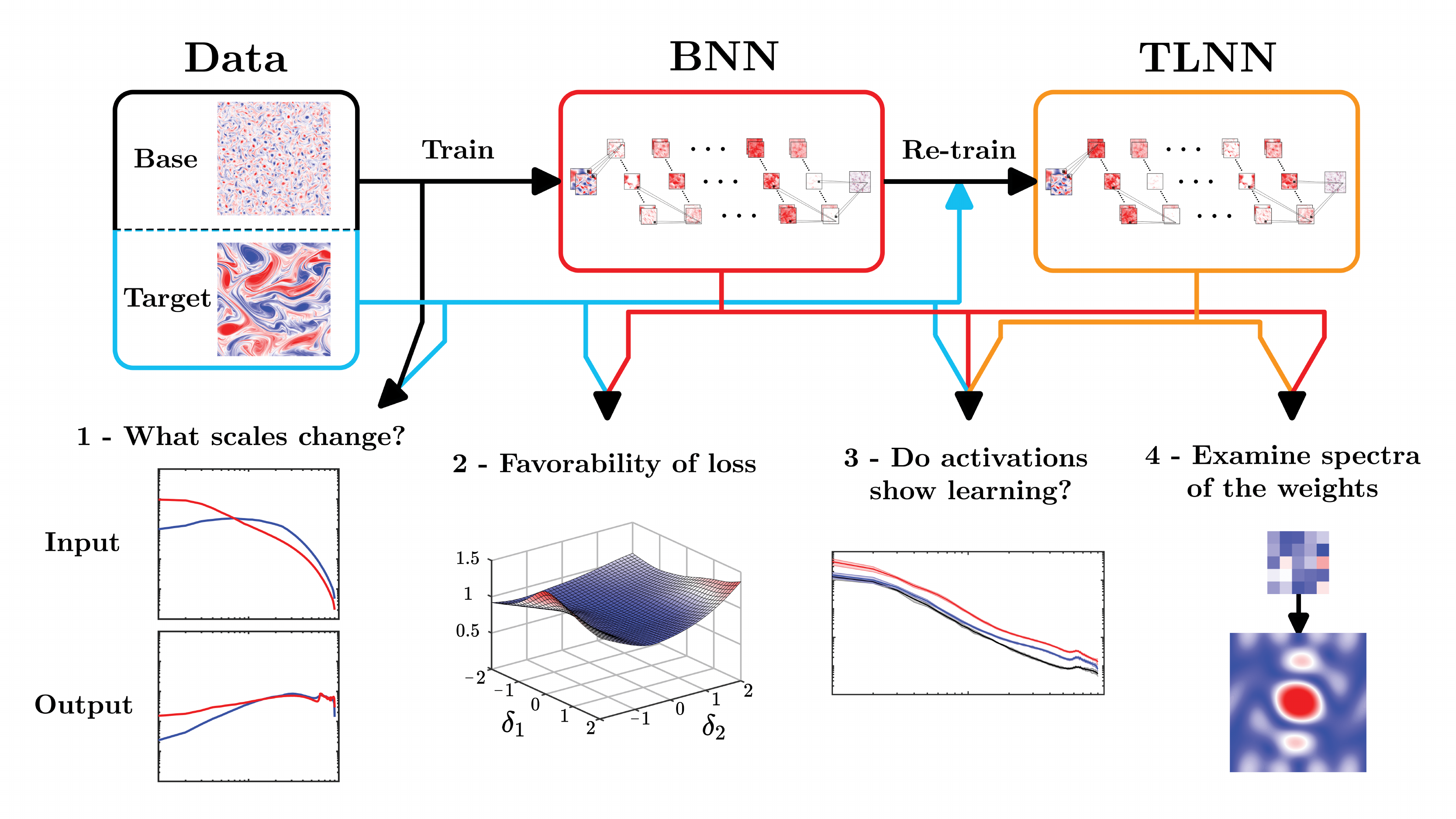}
    \caption{Overview of the framework for guiding and explaining TL onto a new target system. The top row shows the steps of the TL process: acquiring a large amount of training data from the base system and a small amount from the target system, training a BNN$_{base}$ using data from the base system, and re-training it using data from the target system to obtain a TLNN. On the bottom, we present the analyses involved in this framework, listed (left to right) in the order of when they should be used. The arrows indicate what is needed from each step of the TL process and the corresponding analyses. Here, the blue line represents data from the target system, the red line represents the trained BNN$_{base}$, and the orange line represents the re-trained TLNN.  }
    \label{fig:Framework}
\end{figure}

The framework is shown schematically in Fig.~\ref{fig:Framework}. Assuming that we have a large number of training samples from the base system, an accurate BNN$_{base}$ already trained on these samples, and a small number of re-training samples from the target system, the framework involves the following steps:\begin{enumerate}
    \item Compare the spectra of the input and output variables from the base and target systems. The 3 cases studied here have shown that the change of spatial scales between the base and target systems, particularly in the output variables, significantly impacts which layers are optimal for re-training.
    \item Compute the loss landscapes of the BNN$_{base}$ with target systems' data as various combinations of layers are chosen for re-training. Re-training layer(s) with favorable landscapes (smooth and small magnitudes) should be the first choices for TL. We further suggest examining the properly clustered weights' spectra of the BNN$_{base}$ to see if they have clear interpretations as spectral filters.
    \item Re-train a TLNN based on the outcome of Step~2. Examine the spectra of the activations from the re-trained layer(s) and the last hidden layer to see if the differences in the spatial scales identified in Step~1 are learned.
    \item  Examine the spectra of the most-changed kernels between BNN$_{base}$ and TLNN. Investigate if the nature of the newly learned kernels (as spectral filters) is consistent with the outcome of Steps~1 and 3 in terms of spatial scales that need to be learned in TL.
\end{enumerate}
Steps 1-2 are non-intrusive, inexpensive analyses that do not require any re-training, and will effectively guide Step~3, replacing expensive and time-consuming trial-and-error with many combinations of re-training layers. Steps 3-4 provide an explanation for what is learned in TL and act to validate decisions made based on steps 1-2. \\

There are a few points about this framework that need to be further clarified. In general, turbulent flows have universal behavior in their smallest scales \cite{kolmogorov1941local,pope2001turbulent} and vary in large scales due to forcing and geometry. This might seem to suggest that TL will always need to learn changes in large scales between a base and a target turbulent flow. This is not necessarily true, as even in Cases 1-2 here, in which the base and target flows are different in forcing and $Re$ number, there are differences in small-scales of $\Pi$ too. Furthermore, in the broader applications of TL (e.g., in blending different datasets) and beyond just single-physics turbulent flows, there might be differences between the base and target systems at any scales. Step~1 is intended to identify these differences. \\

We also emphasize that currently there is no complete theoretical understanding of which layers of a CNN are better in learning what spatial scales. Our findings for Cases 1-3 and some other studies \cite{rahaman2019spectral,neyshabur_what_2021} in the ML community suggest that the shallower layers are better in learning large scales. If further work confirms this behavior for a variety of systems and CNN architectures, then Steps~1-2 together would be able to even better guide TL in terms of the best layer(s) to re-train. \\

It should be noted that in more complex, an-isotropic, in-homogeneous systems (e.g., channel flows or ocean circulations), spectral analysis using other basis functions, such as Chebyshev or wavelets~\cite{ha2021adaptive,bruna2013spectral}, might be needed. \\

Finally, we point out that a number of recent studies have proposed improving out-of-distribution generalization via incorporating physics constraints into NNs \cite[e.g.,][]{beucler2021climate,kashinath2021physics}. While promising for specific applications, such an approach requires the existence of a physical constraint that is universal (e.g., a scaling law), otherwise, it would deteriorate the performance of the NN. However, the availability of such constraints are very limited. In contrast, TL provides a flexible framework that beyond improving out-of-distribution generalization, is also broadly useful to blend disparate datasets for training, an important application on its own. \\

To summarize, here we have presented the first full explanation of the physics learned in TL for multi-scale, nonlinear dynamical systems, and a novel general framework to guide and explain TL for such systems. This framework will benefit a broad range of applications in areas such as turbulence modeling and weather/climate prediction. Climate change modeling, which deals with an inherently non-stationary system and also involves combining various observational and model datasets, is an application that particularly needs TL, and can benefit from the framework proposed here.


\appendix
\section{Methods and Data}

\subsection{Numerical solvers for DNS and LES}\label{numerical simulation}
We have performed DNS for all 6 systems used in this study (see Table~\ref{tab:table_setups} and Section~\ref{sec:Systems}). In DNS, Eqs.~(\ref{eq:NS1})-(\ref{eq:NS2}) are solved using a Fourier-Fourier pseudo-spectral solver with $N_{\rm{DNS}}$ collocation grid points and second-order Adams-Bashforth and Crank-Nicolson time-integration schemes with time step $\Delta t_{\rm{DNS}}$ for the advection and viscous terms, respectively. See Guan~et al.~\cite{guan_stable_2021,guan2022learning} for more details on the solvers and these simulations. For the base system in Case~1 (decaying 2D turbulence), following earlier studies \citep{maulik2019subgrid,guan_stable_2021}, the flow is initialized randomly using a vorticity field ($\omega_{ic}$) with a prescribed power spectrum. Snapshots of $(\omega,\psi)$ in this system are obtained from $50-200\tau$, where $\tau$ is the initial eddy-turn-over time: $\tau = 1/\text{max}(\omega_{\text{ic}})$. For the other 5 systems (forced 2D turbulence), once the randomly initialized flow reaches statistical equilibrium after a long-term spin-up, we take sequential snapshots of $(\omega,\psi)$ that are $1000\Delta t_\mathrm{DNS}$ apart, in order to reduce the correlation between samples. We use the filtered and coarse-grained DNS data, referred to as FDNS data (Section~\ref{sec:filtering}), for training the CNN-based data-driven closures for $\Pi$ and for testing their {\it a priori} (offline) and {\it a posteriori} (online) performance.   \\

For LES, we solve Eqs.~(\ref{eq:FNS1})-(\ref{eq:FNS2}) employing the same numerical solver used for DNS, but with coarser grid resolutions ($N_\mathrm{LES}=128<N_{\rm{DNS}}$) and larger time steps ($\Delta t_{\mathrm{LES}}= 10\Delta t_{\mathrm{DNS}}$). To represent $\Pi$, a CNN-based closure that is trained on FDNS data (see Section~\ref{sec:CNN}) is coupled to the LES solver.


\subsection{Filtering and coarse-graining: LES equations and FDNS data} \label{sec:filtering}
Filtering Eqs.~\eqref{eq:NS1}-\eqref{eq:NS2} yields the governing equations for LES~\cite{pope2001turbulent,sagaut2006large,guan2022learning}:
\begin{eqnarray}
\frac{\partial \overline{\omega}}{\partial t} + \mathcal{N}(\overline{\omega},\overline{\psi})&=&\frac{1}{Re}\nabla^2\overline{\omega}-\overline{f}-r\overline{\omega}+\underbrace{\mathcal{N}(\overline{\omega},\overline{\psi}) - \overline{\mathcal{N}({\omega},{\psi})}}_{\Pi}\label{eq:FNS1},\\
\nabla^2\overline{\psi} &=& -\overline{\omega}\label{eq:FNS2}.
\end{eqnarray}
In LES, only the large-scale structures ($\overline{\psi}$ and $\overline{\omega}$) are resolved using a coarser grid resolution (compared to DNS). The effects of the structures smaller than the grid spacing are included in the unclosed SGS term $\Pi$, which requires a closure in terms of the resolved flow, ($\overline{\psi},\overline{\omega}$). \\

To obtain the FDNS data, we use the DNS snapshots of (${\psi},{\omega}$), which are of size $N_{\rm{DNS}} \times N_{\rm{DNS}}$, to compute snapshots of $\overline{\psi}$, $\overline{\omega}$, and $\Pi$ (defined in Eq.~\eqref{eq:FNS1}), where $\overline{(\cdot)}$ represents filtering and coarse-graining. The latter is needed to compute these variables on the LES grid (size: $N_{\rm{LES}} \times N_{\rm{LES}}$). Here, we use a Gaussian filter and then sharp spectral cutoff coarse-graining~\citep{guan_stable_2021,guan2022learning}. For each system, the FDNS dataset is divided into completely independent training, validation, and testing sets~\citep{guan_stable_2021,guan2022learning}.

\subsection{Cases 1-3: Base and target systems}
\label{sec:Systems}
By changing $Re$, $r$, $m_f$, and $n_f$, we have created 6 distinct systems of 2D turbulence, which are grouped into 3 cases, each with a base and a target system (Table~\ref{tab:table_setups}). Snapshots of $\omega$ and $\Pi$ as well as the spectra of $\bar{\omega}$, $\Pi$, and KE of these systems are shown in Fig.~\ref{fig:figure_1} to demonstrate the rich variety of fluid flow characteristics among these systems, particularly between each case's base and target systems. Case~1 involves TL from decaying to forced 2D turbulence. From the ${\omega}$ and $\Pi$ snapshots as well as their spectra shown in Fig.~\ref{fig:figure_1}, it is clear that the two systems are different  at both the large and small scales. The significant differences across all scales make this case the most challenging one, and result in the largest generalization gap as discussed in the main text. \\

Case~2 involves TL between two forced 2D turbulence systems: the base system has $Re=10^3$ and the target system has a $100 \times$ higher Reynolds number ($Re = 10^5$), making this the largest extrapolation in $Re$ using TL ever reported, to the best of our knowledge. The increase in $Re$ adds more small-scale features in $\bar{\omega}$ (see the spectrum), and changes the spectrum of $\Pi$ in both large and small scales. Case~3 involves decreasing the forcing wavenumbers of the system. Here, the base system has $m_f=n_f = 25$ while the target system has $m_f=n_f = 4$. This decrease in forcing wavenumbers, as expected, results in more (less) large-scale (small-scale) structures in the resolved flow; see the spectra of $\bar{\omega}$ and KE. Furthermore, more large-scale structures appear in $\Pi$ without any noticeable change in the small-scale structures (see the power spectrum of $\Pi$).

\begin{table}[tbh]
    \centering
    \begin{tabular}{|c|c|c|c|c|c|c|}
    \hline
    System  &  $Re$ & $m_f$   & $n_f$  & $r$  & $N_{\rm{DNS}}$ & $N_{\rm{LES}}$\\
    \hline \hline
    Base (Case 1)  &  $3.2 \times 10^4$  & 0 & 0 & 0 & 2048 & 128 \\
    \hline
    Target (Case 1)  &  $1 \times 10^4$  & 4 & 0 & 0.1 & 1024 & 128 \\
    \hline \hline
    Base (Case 2) &  $1 \times 10^3$  & 4 & 0 & 0.1 & 512 & 128 \\
    \hline
    Target (Case 2) &  $1 \times 10^5$  & 4 & 0 & 0.1 & 2048 & 128\\
    \hline \hline
    Base (Case 3) &  $2 \times 10^4$  & 25 & 25 & 0.1 & 1024 & 128  \\
    \hline
    Target (Case 3) &  $2 \times 10^4$  & 4 & 4 & 0.1 & 1024 & 128\\
    \hline

    \end{tabular}
    \caption{Physical and numerical parameters for the 6 different systems, which are divided into 3 cases, each with a base and a target system. See Fig.~\ref{fig:figure_1} for snapshots and some of the statistical properties of these distinctly different flows.}
    \label{tab:table_setups}
\end{table}

\subsection{Convolutional neural network (CNN) and transfer learning (TL)}\label{sec:CNN}
Building on the success of our earlier work \cite{guan_stable_2021,guan2022learning}, to develop non-local data-driven SGS closure for each system, we train a CNN with input $\mathbf{u} = \left(\bar{\omega}(x,y), \bar{\psi}(x,y) \right)$ to predict $\Pi(x,y)$ (output). These CNNs are built entirely from 11 sequential convolution layers, $9$ of which are hidden layers each with $64^2$ kernels of size $5\times5$ (note that these numbers are hyperparameters that have been optimized for this application \cite{guan_stable_2021,guan2022learning}). The outputs of a convolutional layer are called activations. For channel $j$, of layer $\ell$, the equation for activation $g_\ell^j \in \mathbb{R}^{N_{\rm{LES}} \times N_{\rm{LES}}}$ is:
\begin{equation}
\label{eq:activations}
    g_\ell^j(\mathbf{u}) = \sigma\left(\sum_\beta \left( W_\ell^{\beta,j}\circledast g_{\ell-1}^\beta(\mathbf{u})\right)+b_\ell^j\right).
\end{equation}
Note that $N_{\rm{LES}}=128$ for all systems (Table~\ref{tab:table_setups}). Here, $\circledast$ represents spatial convolution and $\sigma(\cdot)=\max{(0,\cdot)}$ is the ReLU activation function (which is not present for the linear output layer, $\ell=11$). $W^{\beta,j}_\ell \in \mathbb{R}^{5 \times 5}$ is the weight matrix of a convolution kernel, and $b^j_\ell \in \mathbb{R}^{128 \times 128}$ is the regression bias, a constant matrix. We have $\beta \in \{1, 2 \cdots 64\}$ and $j \in \{1, 2 \cdots 64\}$ for all layers with two exceptions: in the input layer ($\ell = 1$) $\beta \in \{1, 2\}$, and in the output layer ($\ell = 11$), $j=1$, as the output is a single channel. The kernels' weights and biases together constitute the NN's trainable parameters, which we collectively refer to as $\theta \in \mathbb{R}^p$. Note that $g_{in}=g_0=\mathbf{u}$ and $g_{out}=g_{11}=\Pi$.    \\


A visualization of these networks as well as examples of activations in the hidden layers are presented in Fig.~\ref{fig:activations}. An important distinction between these CNNs and traditional CNNs is that these do not include any max-pooling layers or dense layers such that they maintain the dimension of the input through all layers and channels in the network. Our earlier work and a few other studies have found such an architecture to lead to more accurate CNNs for SGS closures~\cite{zanna2020data,guan_stable_2021,guan2022learning}.  \\

We train these CNNs using the Adam optimizer and a mean-squared-error (MSE) loss function $\mathcal{L}$. For BNNs, all their trainable parameters $\theta$ are {\it randomly} initialized, and each CNN is trained for 100 epochs using $M_{tr}=2000$ samples from the training set of the base system\footnote{While $M_{tr}=2000$ might seem like a small number of training samples, we are in fact here using a {\it big} training set, because these samples are chosen far apart to be weakly correlated, requiring a long DNS dataset (two million $\Delta t_{\rm{DNS}}$). See Guan et al.~\cite{guan2022learning} for further discussions about the big versus small training sets.}. Note that even when we use $M_{tr}$ samples from the {\it training} set of the {\it target} system to train a CNN, we still call it a  ``BNN'' for convenience (e.g., in Fig.~\ref{fig:figure_1}). Subscripts on BNNs clearly indicate which system provided the $M_{tr}$ training samples. \\

To perform TL from a BNN, the weights and biases of the TLNN are initialized with those of the BNN. The layers to re-train are selected (trainable layers) and the remaining weights/biases are frozen (non-trainable layers). The TLNN is then re-trained using standard backpropagation and the same MSE loss function with $M_{tr}/10$ samples from the training set of the {\it target system}, updating the weights and biases of the trainable layers. The re-training continues until the loss plateaus (for TL, this happens at around $50$ epochs). Note that based on offline metrics such as the correlation coefficients for $\Pi$, we have not found any need for adjusting the hyperparameters such as the learning rate or adding additional layers between training a BNN and TLNN.


\subsection{Spectral analysis of CNNs}
\label{sec:spec_analysis}
The Fourier transform operator $\mathcal{F}$ is defined as

\begin{equation}
\label{eq:fft_normal}
    \hat{\cdot} = \mathcal{F} \left(\cdot  \right), \quad \mathcal{F}: ~ \mathbb{R}^{128\times 128} \longmapsto \mathbb{C}^{128\times 128}.
\end{equation}
To represent convolution as an operation in the spectral space, we first note that we can extend each kernel $W_\ell^{\beta,j}\in \mathbb{R}^{5\times 5}$ to the full domain of the input by padding it with zeros, as done in practice for faster training \cite{mathieu2013fast}, to obtain $\widetilde{W}_\ell^{\beta,j} \in \mathbb{R}^{128\times 128}$. Then, the convolution theorem yields
\begin{equation}
\label{eq:fft_weights}
     W_\ell^{\beta,j}\circledast g_{\ell-1}^\beta = \mathcal{F}^{-1}\left(\hat{\widetilde{W}}_\ell^{\beta,j}\odot \hat{g}_{\ell-1}^\beta\right),
\end{equation}
where $\odot$ is element-wise multiplication. \\

Next, we define linear activation $h_\ell^j$, which contains all the linear operations in Eq.~(\ref{eq:activations}):
\begin{equation}
    h_\ell^j = \sum_\beta \left( W_\ell^{\beta,j}\circledast g_{\ell-1}^\beta\right)+b_\ell^j.
    \label{eq:act_linear}
\end{equation}
Despite the nonlinearity of Eq.~(\ref{eq:activations}) due to the ReLU function, its Fourier transform can be written analytically. Using Eqs.~(\ref{eq:fft_weights}) and \eqref{eq:act_linear} and the linearity of the Fourier transform we obtain
\begin{equation}
    \hat{g}_{\ell}^j = \sum_{\alpha} \left(e^{-i(k_{x}x_{\alpha}+k_{y}y_{\alpha})}\right) \circledast \hat{h}_{\ell}^j = \sum_{\alpha} \left(e^{-i(k_{x}x_{\alpha}+k_{y}y_{\alpha})}\right) \circledast \left\{\sum_\beta  \left(\hat{\widetilde{W}}_\ell^{\beta,j}\odot \hat{g}_{\ell-1}^\beta\right)+\hat{b}_\ell^j \right\},
    \label{eq:gspectra}
\end{equation}
where $(x_\alpha,y_\alpha) \in \left\{(x,y)~|~h_\ell^j(x,y) > 0 \right\}$ and $i = \sqrt{-1}$. The term with sum over $\alpha$ is a result of the ReLU function and involves summing over grid points where $h_\ell^j> 0$. Note that $e^{-i(k_{x}x_{\alpha}+k_{y}y_{\alpha})}$ is the Fourier transform of the delta function at a point $(x_\alpha,y_\alpha)$. Also note that ${b}_\ell^j$ is a constant matrix, therefore, $\hat{b}_\ell^j$ is only non-zero  at $k_x=k_y=0$ (and is real). \\

Equation~\eqref{eq:gspectra} shows that the spectrum of $\hat{g}_{\ell}^j$ depends on the spectrum of $\hat{g}_{\ell-1}^j$, the spectra of the weights $\hat{\widetilde{W}}_\ell^{\beta,j}$ (and constant biases $\hat{b}_\ell^j$), and where $h_{\ell}^j>0$ in the physical (grid) space. With TL, the weights and biases are updated, which changes their spectra as well as where $h_{\ell}^j>0$. Understanding the full effects of all these changes on $\hat{g}_{\ell}^j$ is challenging. In Supplementary Figure~1, we have examined the spectra of activations of layers 2 and 10 from BNN$_{base}$, TLNN$^2$, and TLNN$^{10}$ before and after applying the ReLU activation function (i.e., compare the spectra of $\hat{h}^j_\ell$ and $\hat{g}^j_\ell$). This analysis shows that in all 3 cases, linear changes due to updating $\hat{h}^j_\ell$ substantially alter the spectra of the activations while nonlinear changes only play a significant role in Case~1. These results and Eq.~\eqref{eq:gspectra} suggest that a deeper insight into TL might be obtained by investigating  $\hat{\widetilde{W}}_\ell^{\beta,j}$ and how they change from BNN$_{base}$ to TLNN.


\subsection{Calculating the loss landscape}
\label{sec:loss_land}
Let's represent a CNN with input $\mathbf{u}$ and trainable parameters $\theta$ as $\mathcal{C} \left(\mathbf{u},\theta \right)$. The MSE loss function of this CNN is a function of the output: $\mathcal{L}\left( \mathcal{C} \right)$. The concept of loss landscape (of $\mathcal{L}$) has received much attention in recent years and is widely used to study the {\it training phase} of NNs  \citep{li_visualizing_2018,krishnapriyan2021characterizing, mojgani2022lagrangian}. Below, leveraging recent work in theoretical ML \cite{neyshabur_what_2021}, we compute the loss landscape to study the {\it re-training phase} of NNs in order to gain insight into TL.  \\

Suppose that $\theta_\ell \in \mathbb{R}^{p}$ are all the trainable parameters of a BNN$_{base}$ from all layers $\ell$. We define $\theta^{*}_L \in \mathbb{R}^{p^{*}}$ as the subset of parameters that are updated in TL, i.e., the weights and biases of the re-trained layer(s), $L$. Next, we follow two methodologies for constructing loss landscapes. In the first method, we follow Li~et~al.~\cite{li_visualizing_2018} and select two random direction vectors $v_1,v_2 \in \mathcal{R}^{p*}$ and normalize them with the 2-norm of $\theta^{*}$. In the second method, we follow \cite{yao2020pyhessian} and find the eigenvectors of the Hessian of $\mathcal{L}\left( \mathcal{C} \right)$ computed with respect to $\theta^{*}_L$. The first two eigenvectors with largest positive eigenvalues are chosen as $v_1$ and $v_2$.

Next, in both methods, we perturb $\theta^{*}$ along directions $v_1$ and $v_2$ by amplitudes $\delta_1$ and $\delta_2$, respectively ($\delta_1,\delta_2 \in [-2,2]$ for method~1, $[-1,1]$ for method~2). Finally, we compute
\begin{eqnarray}
\mathcal{L}_{(\delta_1,\delta_2)}= \mathcal{L}\left( \mathcal{C} \left(\mathbf{u}_{target},\left[\theta_{\ell \ne L} \; \; \theta^*_L+\delta_{1} v_1 + \delta_{2} v_{2}\right]\right) \right)
\label{eq:landscape}
\end{eqnarray}
to generate a 2D approximation of the loss landscape and plot the surface as a function of $\delta_1$ and $\delta_2$. Note that the input $\mathbf{u}$ is from the {\it target} system. Loss landscapes from the first (second) method are shown in Fig.~\ref{fig:Loss} (Supplementary Fig.~4). \\

{In the context of TL, the shape of the loss landscape indicates how receptive the re-training layers, $L$, are to change for the new re-training samples from the target system. In practice, a \textit{shallow}, \textit{convex} landscape suggests that the network is in a favorable region of parameter space, and gradient descent will easily converge. Deviations from this in the form of pathological non-convexities or extremely large loss magnitudes can cause problems during training and prevent the network from converging to a useful optimum.} See Li~et~al.~\citep{li_visualizing_2018} and Krishnapriyan~et~al.~\citep{krishnapriyan2021characterizing} for further discussions on the interpretation of loss landscapes for the common application where, in Eq.~\eqref{eq:landscape}$, \mathbf{u}$ is from the base system and $\theta^*$ represent parameters still changing during the epochs of training.

\section*{Supplementary Information}
The supplementary information can be found \href{https://drive.google.com/file/d/1Wp2RR5azECUpR8ehiHKCunqiJmk7rXD3/view?usp=sharing}{here}.

\section*{Acknowledgments}
{We thank Laure Zanna for insightful discussions}. This work was supported by an award from the ONR Young Investigator Program (N00014-20-1-2722), a grant from the NSF CSSI program (OAC-2005123), and by the generosity of Eric and Wendy Schmidt by recommendation of the Schmidt Futures program. Computational resources were provided by NSF XSEDE (allocation ATM170020) and NCAR’s CISL (allocation URIC0004). Our codes and data are available at~\url{https://github.com/envfluids/TL_for_SGS_Models}.


\pagebreak
\setlength{\bibsep}{2.6pt plus 1ex}
\begin{spacing}{.01}
	\small
\bibliographystyle{unsrt}
\bibliography{Bib3}

\begin{thebibliography}{10}

\bibitem{schneider2017earth}
Tapio Schneider, Shiwei Lan, Andrew Stuart, and Joao Teixeira.
\newblock Earth system modeling 2.0: A blueprint for models that learn from
  observations and targeted high-resolution simulations.
\newblock {\em Geophysical Research Letters}, 44(24):12--396, 2017.

\bibitem{brenowitz2018prognostic}
Noah~D Brenowitz and Christopher~S Bretherton.
\newblock Prognostic validation of a neural network unified physics
  parameterization.
\newblock {\em Geophysical Research Letters}, 45(12):6289--6298, 2018.

\bibitem{rasp2018deep}
Stephan Rasp, Michael~S Pritchard, and Pierre Gentine.
\newblock Deep learning to represent subgrid processes in climate models.
\newblock {\em Proceedings of the National Academy of Sciences},
  115(39):9684--9689, 2018.

\bibitem{bolton2019applications}
Thomas Bolton and Laure Zanna.
\newblock Applications of deep learning to ocean data inference and subgrid
  parameterization.
\newblock {\em Journal of Advances in Modeling Earth Systems}, 11(1):376--399,
  2019.

\bibitem{beck2019deep}
Andrea Beck, David Flad, and Claus-Dieter Munz.
\newblock Deep neural networks for data-driven {LES} closure models.
\newblock {\em Journal of Computational Physics}, 398:108910, 2019.

\bibitem{ham_deep_2019}
Yoo-Geun Ham, Jeong-Hwan Kim, and Jing-Jia Luo.
\newblock Deep learning for multi-year {ENSO} forecasts.
\newblock {\em Nature}, 573(7775):568--572, September 2019.

\bibitem{raissi2019physics}
Maziar Raissi, Paris Perdikaris, and George~E Karniadakis.
\newblock Physics-informed neural networks: A deep learning framework for
  solving forward and inverse problems involving nonlinear partial differential
  equations.
\newblock {\em Journal of Computational physics}, 378:686--707, 2019.

\bibitem{weyn2020improving}
Jonathan~A Weyn, Dale~R Durran, and Rich Caruana.
\newblock Improving data-driven global weather prediction using deep
  convolutional neural networks on a cubed sphere.
\newblock {\em Journal of Advances in Modeling Earth Systems},
  12(9):e2020MS002109, 2020.

\bibitem{brunton_machine_2020}
Steven~L Brunton, Bernd~R Noack, and Petros Koumoutsakos.
\newblock Machine learning for fluid mechanics.
\newblock {\em Annual Review of Fluid Mechanics}, 52:477--508, 2020.

\bibitem{yuval2020stable}
Janni Yuval and Paul~A O’Gorman.
\newblock Stable machine-learning parameterization of subgrid processes for
  climate modeling at a range of resolutions.
\newblock {\em Nature communications}, 11(1):1--10, 2020.

\bibitem{kochkov2021machine}
Dmitrii Kochkov, Jamie~A Smith, Ayya Alieva, Qing Wang, Michael~P Brenner, and
  Stephan Hoyer.
\newblock Machine learning--accelerated computational fluid dynamics.
\newblock {\em Proceedings of the National Academy of Sciences}, 118(21), 2021.

\bibitem{novati2021automating}
Guido Novati, Hugues~Lascombes de~Laroussilhe, and Petros Koumoutsakos.
\newblock Automating turbulence modelling by multi-agent reinforcement
  learning.
\newblock {\em Nature Machine Intelligence}, 3(1):87--96, 2021.

\bibitem{pathak2022fourcastnet}
Jaideep Pathak, Shashank Subramanian, Peter Harrington, Sanjeev Raja, Ashesh
  Chattopadhyay, Morteza Mardani, Thorsten Kurth, David Hall, Zongyi Li, Kamyar
  Azizzadenesheli, Pedram Hassanzadeh, Karthik Kashinath, and Animashree
  Anandkumar.
\newblock {FourCastNet}: A global data-driven high-resolution weather model
  using adaptive {F}ourier neural operators.
\newblock {\em arXiv preprint arXiv:2202.11214}, 2022.

\bibitem{yosinski_how_2014}
Jason Yosinski, Jeff Clune, Yoshua Bengio, and Hod Lipson.
\newblock How transferable are features in deep neural networks?
\newblock {\em arXiv:1411.1792 [cs]}, November 2014.
\newblock arXiv: 1411.1792.

\bibitem{nagarajan2020understanding}
Vaishnavh Nagarajan, Anders Andreassen, and Behnam Neyshabur.
\newblock Understanding the failure modes of out-of-distribution
  generalization.
\newblock {\em arXiv preprint arXiv:2010.15775}, 2020.

\bibitem{chattopadhyay_datadriven_2020}
Ashesh Chattopadhyay, Adam Subel, and Pedram Hassanzadeh.
\newblock Data-driven super-parameterization using deep learning:
  {E}xperimentation with multiscale {L}orenz 96 systems and transfer learning.
\newblock {\em Journal of Advances in Modeling Earth Systems},
  12(11):e2020MS002084, 2020.

\bibitem{beucler2021enforcing}
Tom Beucler, Michael Pritchard, Stephan Rasp, Jordan Ott, Pierre Baldi, and
  Pierre Gentine.
\newblock Enforcing analytic constraints in neural networks emulating physical
  systems.
\newblock {\em Physical Review Letters}, 126(9):098302, 2021.

\bibitem{subel_data-driven_2021}
Adam Subel, Ashesh Chattopadhyay, Yifei Guan, and Pedram Hassanzadeh.
\newblock Data-driven subgrid-scale modeling of forced {Burgers} turbulence
  using deep learning with generalization to higher {Reynolds} numbers via
  transfer learning.
\newblock {\em Physics of Fluids}, 33(3):031702, March 2021.

\bibitem{chung2021interpretable}
Wai~Tong Chung, Aashwin~Ananda Mishra, and Matthias Ihme.
\newblock Interpretable data-driven methods for subgrid-scale closure in {LES}
  for transcritical {LOX}/{GCH}4 combustion.
\newblock {\em Combustion and Flame}, page 111758, 2021.

\bibitem{frezat2021physical}
Hugo Frezat, Guillaume Balarac, Julien Le~Sommer, Ronan Fablet, and Redouane
  Lguensat.
\newblock Physical invariance in neural networks for subgrid-scale scalar flux
  modeling.
\newblock {\em Physical Review Fluids}, 6(2):024607, 2021.

\bibitem{taghizadeh2020turbulence}
Salar Taghizadeh, Freddie~D Witherden, and Sharath~S Girimaji.
\newblock Turbulence closure modeling with data-driven techniques: physical
  compatibility and consistency considerations.
\newblock {\em New Journal of Physics}, 22(9):093023, 2020.

\bibitem{guan_stable_2021}
Yifei Guan, Ashesh Chattopadhyay, Adam Subel, and Pedram Hassanzadeh.
\newblock Stable a posteriori {LES} of {2D} turbulence using convolutional
  neural networks: {Backscattering} analysis and generalization to higher {Re}
  via transfer learning.
\newblock {\em Journal of Computational Physics (in press)}, February 2021.
\newblock arXiv: 2102.11400.

\bibitem{tan2018survey}
Chuanqi Tan, Fuchun Sun, Tao Kong, Wenchang Zhang, Chao Yang, and Chunfang Liu.
\newblock A survey on deep transfer learning.
\newblock In {\em International conference on artificial neural networks},
  pages 270--279. Springer, 2018.

\bibitem{zhuang2020comprehensive}
Fuzhen Zhuang, Zhiyuan Qi, Keyu Duan, Dongbo Xi, Yongchun Zhu, Hengshu Zhu, Hui
  Xiong, and Qing He.
\newblock A comprehensive survey on transfer learning.
\newblock {\em Proceedings of the IEEE}, 109(1):43--76, 2020.

\bibitem{inubushi2020transfer}
Masanobu Inubushi and Susumu Goto.
\newblock Transfer learning for nonlinear dynamics and its application to fluid
  turbulence.
\newblock {\em Physical Review E}, 102(4):043301, 2020.

\bibitem{guastoni2021convolutional}
Luca Guastoni, Alejandro G{\"u}emes, Andrea Ianiro, Stefano Discetti, Philipp
  Schlatter, Hossein Azizpour, and Ricardo Vinuesa.
\newblock Convolutional-network models to predict wall-bounded turbulence from
  wall quantities.
\newblock {\em Journal of Fluid Mechanics}, 928, 2021.

\bibitem{yousif2021high}
Mustafa~Z Yousif, Linqi Yu, and Hee-Chang Lim.
\newblock High-fidelity reconstruction of turbulent flow from spatially limited
  data using enhanced super-resolution generative adversarial network.
\newblock {\em Physics of Fluids}, 33(12):125119, 2021.

\bibitem{rasp_datadriven_2021}
Stephan Rasp and Nils Thuerey.
\newblock Data-driven medium-range weather prediction with a {R}es{N}et
  pretrained on climate simulations: {A} new model for weatherbench.
\newblock {\em Journal of Advances in Modeling Earth Systems},
  13(2):e2020MS002405, 2021.

\bibitem{mondal_transfer_2021}
Sudeepta Mondal, Ashesh Chattopadhyay, Achintya Mukhopadhyay, and Asok Ray.
\newblock Transfer learning of deep neural networks for predicting
  thermoacoustic instabilities in combustion systems.
\newblock {\em Energy and AI}, 5:100085, September 2021.

\bibitem{chattopadhyay2022long}
Ashesh Chattopadhyay, Jaideep Pathak, Ebrahim Nabizadeh, Wahid Bhimji, and
  Pedram Hassanzadeh.
\newblock Long-term stability and generalization of observationally-constrained
  stochastic data-driven models for geophysical turbulence.
\newblock {\em arXiv preprint arXiv:2205.04601}, 2022.

\bibitem{hu2021deep}
Jie Hu, Bin Weng, Tianqiang Huang, Jianyun Gao, Feng Ye, and Lijun You.
\newblock Deep residual convolutional neural network combining dropout and
  transfer learning for enso forecasting.
\newblock {\em Geophysical Research Letters}, 48(24):e2021GL093531, 2021.

\bibitem{karniadakis2021physics}
George~Em Karniadakis, Ioannis~G Kevrekidis, Lu~Lu, Paris Perdikaris, Sifan
  Wang, and Liu Yang.
\newblock Physics-informed machine learning.
\newblock {\em Nature Reviews Physics}, 3(6):422--440, 2021.

\bibitem{chakraborty2021transfer}
Souvik Chakraborty.
\newblock Transfer learning based multi-fidelity physics informed deep neural
  network.
\newblock {\em Journal of Computational Physics}, 426:109942, 2021.

\bibitem{hussain2018study}
Mahbub Hussain, Jordan~J Bird, and Diego~R Faria.
\newblock A study on cnn transfer learning for image classification.
\newblock In {\em UK Workshop on computational Intelligence}, pages 191--202.
  Springer, 2018.

\bibitem{talo2019application}
Muhammed Talo, Ulas~Baran Baloglu, {\"O}zal Y{\i}ld{\i}r{\i}m, and U~Rajendra
  Acharya.
\newblock Application of deep transfer learning for automated brain abnormality
  classification using mr images.
\newblock {\em Cognitive Systems Research}, 54:176--188, 2019.

\bibitem{zeiler2014visualizing}
Matthew~D Zeiler and Rob Fergus.
\newblock Visualizing and understanding convolutional networks.
\newblock In {\em European conference on computer vision}, pages 818--833.
  Springer, 2014.

\bibitem{maulik2019subgrid}
Romit Maulik, Omer San, Adil Rasheed, and Prakash Vedula.
\newblock Subgrid modelling for two-dimensional turbulence using neural
  networks.
\newblock {\em Journal of Fluid Mechanics}, 858:122--144, 2019.

\bibitem{page2021revealing}
Jacob Page, Michael~P Brenner, and Rich~R Kerswell.
\newblock Revealing the state space of turbulence using machine learning.
\newblock {\em Physical Review Fluids}, 6(3):034402, 2021.

\bibitem{pawar2022frame}
Suraj Pawar, Omer San, Adil Rasheed, and Prakash Vedula.
\newblock Frame invariant neural network closures for kraichnan turbulence.
\newblock {\em arXiv preprint arXiv:2201.02928}, 2022.

\bibitem{guan2022learning}
Yifei Guan, Adam Subel, Ashesh Chattopadhyay, and Pedram Hassanzadeh.
\newblock Learning physics-constrained subgrid-scale closures in the small-data
  regime for stable and accurate les.
\newblock {\em arXiv preprint arXiv:2201.07347}, 2022.

\bibitem{neyshabur_what_2021}
Behnam Neyshabur, Hanie Sedghi, and Chiyuan Zhang.
\newblock What is being transferred in transfer learning?
\newblock {\em arXiv:2008.11687 [cs, stat]}, January 2021.
\newblock arXiv: 2008.11687.

\bibitem{zhuang_comprehensive_2021}
Fuzhen Zhuang, Zhiyuan Qi, Keyu Duan, Dongbo Xi, Yongchun Zhu, Hengshu Zhu, Hui
  Xiong, and Qing He.
\newblock A comprehensive survey on transfer learning.
\newblock {\em Proceedings of the IEEE}, 109(1):43--76, 2020.

\bibitem{chizat2019lazy}
Lenaic Chizat, Edouard Oyallon, and Francis Bach.
\newblock On lazy training in differentiable programming.
\newblock {\em Advances in Neural Information Processing Systems}, 32, 2019.

\bibitem{li_visualizing_2018}
Hao Li, Zheng Xu, Gavin Taylor, Christoph Studer, and Tom Goldstein.
\newblock Visualizing the {Loss} {Landscape} of {Neural} {Nets}.
\newblock {\em arXiv:1712.09913 [cs, stat]}, November 2018.
\newblock arXiv: 1712.09913.

\bibitem{krishnapriyan2021characterizing}
Aditi Krishnapriyan, Amir Gholami, Shandian Zhe, Robert Kirby, and Michael~W
  Mahoney.
\newblock Characterizing possible failure modes in physics-informed neural
  networks.
\newblock {\em Advances in Neural Information Processing Systems}, 34, 2021.

\bibitem{rahaman2019spectral}
Nasim Rahaman, Aristide Baratin, Devansh Arpit, Felix Draxler, Min Lin, Fred
  Hamprecht, Yoshua Bengio, and Aaron Courville.
\newblock On the spectral bias of neural networks.
\newblock In {\em International Conference on Machine Learning}, pages
  5301--5310. PMLR, 2019.

\bibitem{kolmogorov1941local}
Andrey~Nikolaevich Kolmogorov.
\newblock The local structure of turbulence in incompressible viscous fluid for
  very large {R}eynolds numbers.
\newblock {\em Cr Acad. Sci. URSS}, 30:301--305, 1941.

\bibitem{pope2001turbulent}
Stephen~B Pope.
\newblock {\em Turbulent {F}lows}.
\newblock IOP Publishing, 2001.

\bibitem{ha2021adaptive}
Wooseok Ha, Chandan Singh, Francois Lanusse, Srigokul Upadhyayula, and Bin Yu.
\newblock Adaptive wavelet distillation from neural networks through
  interpretations.
\newblock {\em Advances in Neural Information Processing Systems}, 34, 2021.

\bibitem{bruna2013spectral}
Joan Bruna, Wojciech Zaremba, Arthur Szlam, and Yann LeCun.
\newblock Spectral networks and locally connected networks on graphs.
\newblock {\em arXiv preprint arXiv:1312.6203}, 2013.

\bibitem{beucler2021climate}
Tom Beucler, Michael Pritchard, Janni Yuval, Ankitesh Gupta, Liran Peng,
  Stephan Rasp, Fiaz Ahmed, Paul~A O'Gorman, J~David Neelin, Nicholas~J Lutsko,
  et~al.
\newblock Climate-invariant machine learning.
\newblock {\em arXiv preprint arXiv:2112.08440}, 2021.

\bibitem{kashinath2021physics}
K~Kashinath, M~Mustafa, A~Albert, JL~Wu, C~Jiang, S~Esmaeilzadeh,
  K~Azizzadenesheli, R~Wang, A~Chattopadhyay, A~Singh, et~al.
\newblock Physics-informed machine learning: case studies for weather and
  climate modelling.
\newblock {\em Philosophical Transactions of the Royal Society A},
  379(2194):20200093, 2021.

\bibitem{sagaut2006large}
Pierre Sagaut.
\newblock {\em Large eddy simulation for incompressible flows: An
  introduction}.
\newblock Springer Science \& Business Media, 2006.

\bibitem{zanna2020data}
Laure Zanna and Thomas Bolton.
\newblock Data-driven equation discovery of ocean mesoscale closures.
\newblock {\em Geophysical Research Letters}, 47(17):e2020GL088376, 2020.

\bibitem{mathieu2013fast}
Michael Mathieu, Mikael Henaff, and Yann LeCun.
\newblock Fast training of convolutional networks through ffts.
\newblock {\em arXiv preprint arXiv:1312.5851}, 2013.

\bibitem{mojgani2022lagrangian}
Rambod Mojgani, Maciej Balajewicz, and Pedram Hassanzadeh.
\newblock Lagrangian {PINNs}: A causality-conforming solution to failure modes
  of physics-informed neural networks.
\newblock {\em arXiv preprint arXiv:2205.02902}, 2022.

\bibitem{yao2020pyhessian}
Zhewei Yao, Amir Gholami, Kurt Keutzer, and Michael~W Mahoney.
\newblock Pyhessian: Neural networks through the lens of the hessian.
\newblock In {\em 2020 IEEE international conference on big data (Big data)},
  pages 581--590. IEEE, 2020.

\end{thebibliography}
\end{spacing}

\end{document}